\documentclass[onecolumn,preprint]{revtex4-1}
\usepackage{graphicx}
\usepackage{amsmath}
\usepackage{wrapfig}

\newcommand{\boutxx}{\texttt{BOUT++ }}

\begin{document}
\title{Blob dynamics in TORPEX poloidal null configurations}

\author{B W Shanahan}
\author{B D Dudson}
\affiliation{York Plasma Institute, Department of Physics, University of York, Heslington, York YO10 5DD, UK}

\email{bws502@york.ac.uk}

\begin{abstract}
 Three dimensional blob dynamics are simulated in X-point magnetic configurations in the TORPEX device via a non-field-aligned coordinate system, using an isothermal model which evolves density, vorticity, parallel velocity and parallel current density. By modifying the parallel gradient operator to include perpendicular perturbations from poloidal field coils, numerical singularities associated with field aligned coordinates are avoided.  A comparison with a previously developed analytical model~\cite{Avino2016} is performed and an agreement is found with minimal modification. Experimental comparison determines that the null region can cause an acceleration of filaments due to increasing connection length, but this acceleration is small relative to other effects, which we quantify.  Experimental measurements~\cite{Avino2016} are reproduced, and the dominant acceleration mechanism is identified as that of a developing dipole in a moving background. Contributions from increasing connection length close to the null point are a small correction.

\end{abstract}
\maketitle

\section{Introduction}
\label{sec:intro}
Filaments, or blobs, are typically field aligned plasma structures which have been observed in the scrape of layer (SOL) of many magnetically confined plasmas~\cite{Dippolito2011}.  These filaments carry particles into the SOL and therefore play a role in determining the profiles during L-mode and inter-ELM H-mode scenarios. While there have been many investigations into the dynamics of such filaments~\cite{Dippolito2011,Walkden2013, Myra2006}, few if any have studied their behavior near magnetic X-points. Simple magnetic tori such as the TORPEX device~\cite{Fasoli2010} replicate tokamak scrape off layer (SOL) scenarios while allowing straightforward diagnostic access.  While filaments have been studied extensively experimentally within TORPEX~\cite{Furno2008,Podesta2008,Theiler2009}, no theoretical studies have yet explored the dynamics in X-point configurations recently achieved experimentally~\cite{Avino2014,FAvino,Avino2016}.

The fundamental physics of blob propagation is described in detail in~\cite{Dippolito2011} which is as follows. The divergence of the diamagnetic drift (physically, the curvature drift) causes a polarization of the blob, leading to an {$\bf{E} \times B$} velocity in the form of counter-rotating vortices and an outward advection of the blob.  The dynamics of propagating filaments depends on the mechanism for charge dissipation within the blob in order to satisfy quasineutrality, $\nabla \cdot {\bf{j}} = 0$.  If the charge separation caused by diamagnetic drifts is resolved primarily via the parallel current through the sheath, the filament is considered to be sheath-connected~\cite{Dippolito2011, Krasheninnikov2001}. If the connection length to the sheath is too large, or likewise the resistivity too high, charge is dissipated via cross-field currents such as the polarization current~\cite{Krasheninnikov2008,Omotani2015,Easy2014} and the blob is said to be in the inertially limited regime~\cite{Garcia2005}.

In this work filaments are characterized in TORPEX magnetic null point scenarios using three dimensional simulations in \boutxx~\cite{Dudson2009}.  The research presented here focuses on the behavior of filaments as they encounter both open and closed field lines, and how that simulated behavior relates to experimentally observed characteristics.  Recent work~\cite{Avino2016} has sought to experimentally characterize filaments in TORPEX magnetic null configurations.  A significant acceleration of filaments towards the X-point is observed in~\cite{Avino2016}, and an analytical model is developed to explain this acceleration.  In the region of poloidal magnetic nulls, the distance along the field lines between the two lobes of the potential dipole, called the connection length $L_\parallel$, is increased.  This increased connection length is considered to reduce the effect of charge dissipation via parallel currents, and therefore an acceleration is manifested.  Interestingly, a deceleration of the filaments is seen experimentally in the immediate vicinity of the X-point, but this is attributed to dissolution of the blob structure.  Here we simulate filaments in these scenarios and compare simulations with this previously derived analytical model~\cite{Avino2016} in an attempt to further understand the nature of filament propagation in regions of poloidal magnetic nulls.

\subsection{TORPEX null point scenarios}
\begin{figure}[h!]
\includegraphics[width = 15cm]{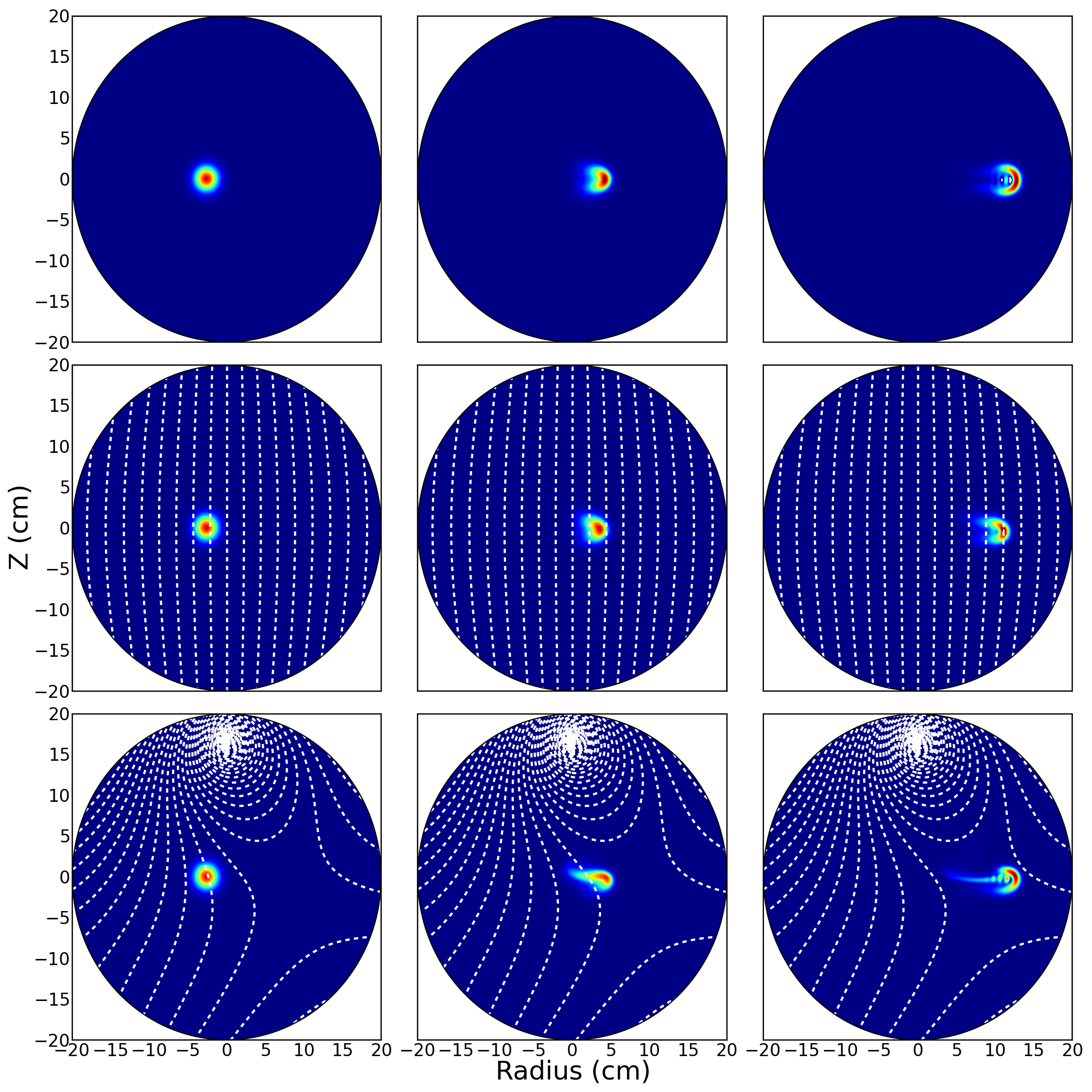}
\caption{\small{Filaments upon initialization (left), 21$\mathrm{\mu}s$ after seeding (middle), and 42$\mathrm{\mu}s$ after seeding (right) in three different TORPEX magnetic geometries; (top) purely toroidal magnetic fields, (middle) a vertical poloidal field, and (bottom) a poloidal magnetic null scenario~\cite{Avino2016}. Poloidal magnetic geometries are indicated by the white, dashed contours.}}
\label{fig:propagation}
\end{figure}

The aim of this work is to explore blob dynamics in the TORPEX simple magnetic torus in X-point geometries~\cite{Avino2014, FAvino, Avino2016}.  Many previous studies of filaments in the TORPEX device~\cite{Riva2016, Easy2014, Halpern2014} utilized a case with a vertical field.  Figure~\ref{fig:propagation} indicates the trajectory of filaments in three different magnetic fields; a purely toroidal field (top), a TORPEX vertical field scenario (middle), and the recently studied poloidal magnetic null scenario (bottom).  

For the vertical and magnetic null scenarios, the magnetic field is calculated based on the coil position and current, which will be discussed further in section~\ref{sec:nummethods}. The TORPEX device has a major radius of 1m, minor radius of 20cm, and a toroidal magnetic field of about 75mT~\cite{Avino2014}.  Filaments in this geometry have been observed experimentally to be toroidally symmetric, and therefore not aligned to the magnetic field~\cite{Avino2016}. The filaments are first considered coherent in experiment 4cm left of the center of the vacuum vessel (r,z = -4,0 cm)~\cite{FAvino, Avino2016}, where r=0 is considered the center of the vacuum vessel.  As such, here we seed toroidally symmetric filaments with an initial peak density of $\mathrm{3\times10^{16}m^{-3}}$ at (r,z) = (-4cm,0cm) with an initial diameter of about 3cm.

\section{Numerical methods and model}
\label{sec:geomodel}

\subsection{Isothermal Model}
\label{sec:torpexmodel}
An isothermal cold-ion fluid model initially constructed for plasma blob studies~\cite{Walkden2013, Angus2012} has been extended for use in X-point scenarios~\cite{Shanahan2014}.  The model is electrostatic and inviscid; the isothermal electron temperature $T_{e0}$ is set to 2.5eV, as this is approximately the measured temperature in the region of filament propagation within TORPEX X-point scenarios~\cite{FAvino}.  The equations which are solved are given as follows in SI units:  
\begin{equation}
\label{eq:density}
\frac{dn}{dt} = (1-\chi)\left[2c_s\rho_s\xi\cdot(\nabla n - {n_0}\nabla\phi) + \nabla_\parallel \frac{J_\parallel}{e} - n_0\nabla_\parallel u_\parallel \right] + \chi\nabla_\parallel^2n
\end{equation}
\begin{equation}
\label{eq:vorticity}
\rho_s^2 n_0\frac{d\Omega}{dt} = 2c_s\rho_s\xi\cdot\nabla n +\nabla_\parallel \frac{J_\parallel}{e}
\end{equation}
\begin{equation}
\label{eq:vparallel}
\frac{du_\parallel}{dt} = -\frac{c_s^2}{n_0}\nabla_\parallel n
\end{equation}
\begin{equation}
J_\parallel = (1-\chi)\left[\frac{\sigma_\parallel T_e }{e{n_0}}(\nabla_\parallel n -  {n_0}\nabla_\parallel\phi)\right]
\end{equation}

Where $\Omega \equiv \nabla_\perp^2 \phi $ is the vorticity, total derivatives are split via $\frac{d}{dt}  = \frac{\partial}{\partial t} + {\bf{u_E}}\cdot\nabla + {\bf{u_\parallel}}\cdot\nabla$, and parallel derivatives are evaluated using $\nabla_\parallel = {\bf{b}} \cdot \nabla $ where {\bf{b}} is the unit vector along the total magnetic field, including the poloidal field. Curvature effects are included via the polarization vector $\xi \equiv \nabla \times \frac{\bf{b}}{B} \sim \frac{1}{BR_c} {\bf{\hat{z}}}$. In the above equations, $\rho_s = \frac{c_s}{\Omega_i}$ is the Bohm gyroradius, and $\sigma_\parallel$ is the parallel (Spitzer~\cite{Cohen1950}) conductivity.  These equations are normalized such that density ($n$) is normalized to typical TORPEX values, $n_0 = 8 \times 10^{15} m^{-3}$, speeds are normalized to the sound speed, and $\phi = \frac{e\Phi}{T_{e0}}$ is the normalized electrostatic plasma potential. 

 Because TORPEX utilizes an in-vessel coil to create the X-point field, the singularity on the coil axis (described in the following section) has been avoided by implementing a penalization scheme~\cite{Isoardi2009}, which utilizes a masking function at the location of the wire such that there are no gradients across the coil cross section. The masking function ($\chi$) has the following form:

\[
\chi =
\begin{cases}
  \hfill 1 \hfill & \text{$ 0 < r < r_c$}\\
  \hfill $$ \chi_0\ ln(r)$$ & \text { $r_c \leq r \leq 1.1 r_c$}\\
  \hfill 0 \hfill & \text{$ r > 1.1 r_c$} \\
\end{cases}
\]

Where $r_c$ is the coil radius, chosen here to be 1cm, and $\chi_0$ is an arbitrary coefficient to determine the smoothness of the masking function.

This model differs from that used in reference~\cite{Walkden2013} in that it incorporates parallel ion free streaming, $u_\parallel$, which could contribute to a radial motion if the field is not strictly toroidal.  In the geometries studied here, however, this effect is found to be small ($10^{-3}$).  Additionally, energy conservation required the restriction that $n$ is considered constant ($n_0$) in terms where it is not differentiated such as the right hand side of Equation~\ref{eq:vparallel}, which is simply a limit of the imposed Boussinesq approximation which assumes that density fluctuations are small: $\nabla \times \left(n \frac{d\nabla_\perp \phi}{dt} \right) \approx n_0 \frac{d}{dt}\nabla_{\perp}^2 \phi$.

\subsection{Numerical Methods} 
\label{sec:nummethods}
 The presence of poloidal magnetic field singularities in the form of O- and X-points in this magnetic topology requires the use of non-field-aligned coordinate systems.  As such, a cylindrical coordinate system defined by the major radius ({\bf{x}}), vertical direction ({\bf{z}}), and toroidal direction ({\bf{y}}) was implemented, and the poloidal magnetic field implemented by prescribing an analytic form for the magnetic vector potential and modifying the parallel gradient operator~\cite{Jackson}: 

\begin{equation}
\label{eq:biotsavart}
{\bf{A(r)} }= \frac{-\mu_0 I}{2 \pi} \ln(r){\bf{\hat{y}}}
\end{equation}
where ${\bf{\hat{y}}}$ is the toroidal direction (parallel to wire).  It is therefore possible to construct an arbitrary magnetic field given the number of turns, current, and location of magnetic coils.  The only difficulty is the infinite magnetic field on axis, which is avoided using a penalization scheme, as described in the previous section.  Our form of the vector potential can therefore be implemented into our simulations as the ${\bf{b}}\cdot\nabla$ operator such that:

\begin{equation}
{\bf{b}}\cdot\nabla f  = \nabla_\parallel f - \left[\frac{A_{ext}}{B},f\right]
\end{equation}

where $A_{ext}$ is the perturbed externally applied vector potential due to the magnetic coils and the square brackets are Poisson brackets which are solved using the Arakawa method~\cite{Arakawa1977}.

The model described in Section~\ref{sec:torpexmodel} is solved in this geometry using a resolution of 1.5mm ($0.36 \rho_s$) in the poloidal plane ({\bf{x}}, {\bf{z}}), and 15.7cm ($36.5\rho_s$) in the toroidal direction ({\bf{y}}).  Time integration was implemented using the implicit time integration solver CVODE, within the SUite of Nonlinear and DIfferential/ALgebraic equation Solvers (SUNDIALS)~\cite{SUNDIALS}. Finally, the Laplacian solver, which calculates potential ($\phi$) from vorticity ($\Omega$), in \boutxx was altered to invert using discrete sine transforms in the {\bf{z}} (vertical) direction, which eliminates the periodicity used in typical Laplacian inversion utilizing Fourier transforms in \boutxx~\cite{Dudson2009}. As the filaments in TORPEX are considered toroidally symmetric and therefore do not reach the sheath, simple Neumann (zero gradient) boundary conditions have been used in the poloidal plane, although presheath boundary conditions~\cite{Loizu2012} have been implemented.

\section{Filament Characterization and Experimental Comparison}

As the model and numerical methods described in the previous section were originally tested in linear geometries~\cite{Shanahan2014}, simulations were performed here to validate the extension of these methods to toroidal geometries and to determine the characteristics of blob propagation within the TORPEX magnetic null point scenarios. Experimental comparison was conducted to investigate the filament acceleration mechanism seen in experiment. The simulations were initialized based on experimental observations~\cite{Avino2016, FurnoPC}; the initial filament diameter, measured as the full width at half maximum, was set to 3cm, and the filaments were seeded at (r,z) = (-4cm,0cm) as this is where filaments are first considered coherent in this TORPEX geometry.  It has been proposed in~\cite{Avino2016} that the poloidal magnetic null region causes an acceleration by increasing the connection length associated with the dipole field.  Here we test this assertion and compare simulations to the previously developed analytical model. 

\subsection{Current analysis}

\begin{figure}[h!]
\includegraphics[width = 15cm]{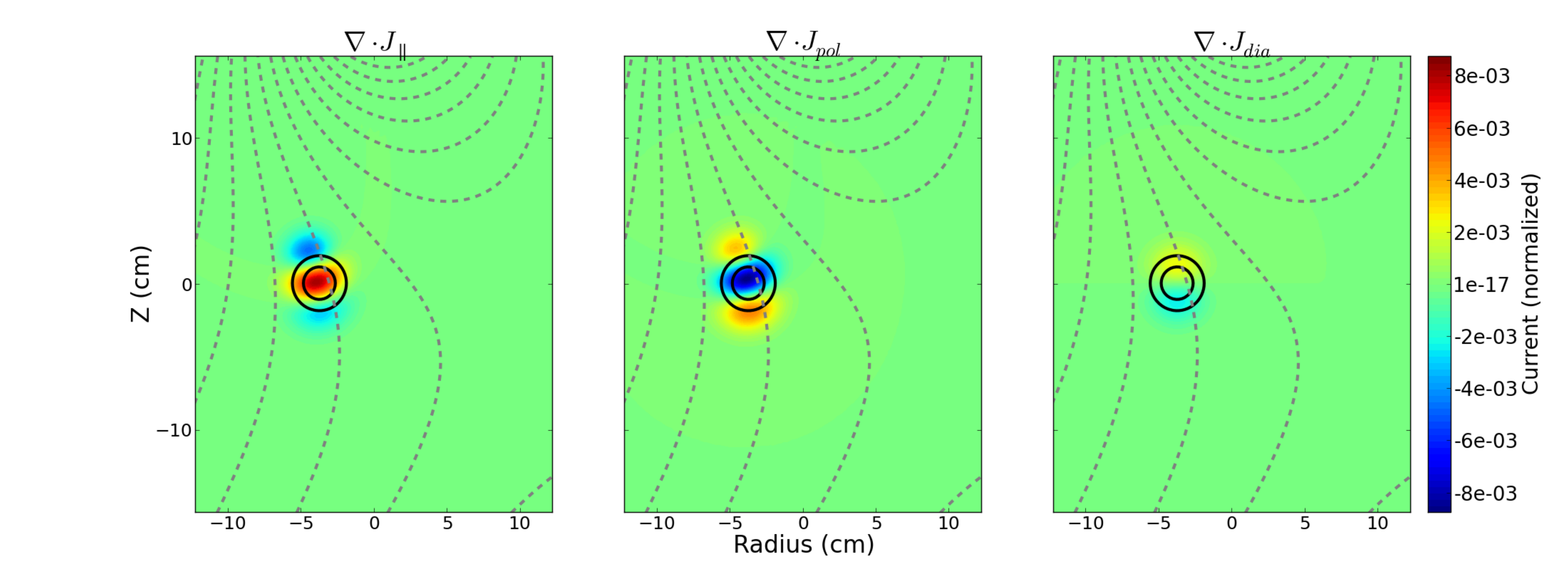}
\caption{\small{The divergence of the various currents within the system (color contour) immediately after initialization.  The blob cross section is shown as the black solid contours, and the poloidal magnetic field is indicated by the grey dashed contours.}}
\label{fig:currentcomparison}
\end{figure}

The currents within filaments determine their propagation.  Typically, filaments are field aligned and therefore can extend to the sheath. This allows the current within the filament to flow through the sheath, although recent work has found that current can still flow to the sheath even if the filament itself does not reach the target~\cite{Easy2014}.  Filaments within this TORPEX configuration have been determined to be toroidally symmetric, however, and therefore the current is expected to be localized within the blob.  As such, we have investigated the currents within the simulated filaments, as shown in Figure~\ref{fig:currentcomparison}.  

From Figure~\ref{fig:currentcomparison} it is apparent that the current is localized to the blob and does not extend to the plasma sheath at the edges of the computational domain.  This localized current is essential to the development of the model in~\cite{Avino2016}, as it is assumed that the charge is dissipated along the field line which connects the two lobes of the potential dipole. 

\subsection{Analytical model comparison}
An analytical model has been previously developed which relies on the assumption of increasing connection length in poloidal magnetic null regions as an acceleration mechanism~\cite{FAvino, Avino2016}. As this model was originally developed in reference~\cite{Avino2016}, it will be referred to here as the ``Avino model''.  In this model, the blob velocity follows a function as shown in Equation~\ref{eq:Avinomodel}:

\begin{equation}
\label{eq:Avinomodel}
v_b = \frac{\delta n}{n} \sqrt{\frac{2a}{R}}c_s\left(\frac{1}{1+A/L_\parallel}\right)
\end{equation}

where:
\begin{equation}
A = \frac{CB^2a^{5/2}\sqrt{2R}}{m_ic_s}
\end{equation}

Here, $C$ is the proportionality coefficient between the plasma conductivity and the plasma density ($C=\sigma/n$), $a$ is the is the radius of the blob, $L_\parallel$ is the parallel connection length, R is the major radius, and $c_s$ is the sound speed.  In the original analysis, the relative perturbation of density, $\delta n/n$, was considered almost constant, and therefore the magnetic field ($B$), which also dictates the parallel connection length $L_\parallel$ across the dipole, is considered the only position-dependent variable.  In the analysis presented here, however, we are able to directly calculate all quantities in Equation~\ref{eq:Avinomodel} from numerical simulations.  


Although the plasma (Spitzer) conductivity~\cite{Cohen1950} is an input to the simulation, it is calculated differently in~\cite{Avino2016}, which defines it as: 

\begin{equation}
\sigma = \frac{ne^2}{m_e \nu_{eH}}
\end{equation}

where $\nu_{eH} = n_n \sigma_{eH} \sqrt{T_e/m_e}$ is the electron-neutral collision frequency.  In these simulations, we have assumed the neutral density $n_n$ is $2\times10^{18}$m$^{-3}$ and a collisional cross section $\sigma_{eH} = 2\times10^{-19}$m$^2$ following the analysis of~\cite{Tawara1990}.  It should be noted, however, that this plasma conductivity only affects the value of $C$, which is used as a free parameter both here and in~\cite{Avino2016} to ensure that the model correctly corresponds with initial measured/simulated filament velocity. As stated previously, an isothermal temperature of 2.5eV was assumed.  

The blob size $a$ can be calculated as half the distance between the maximum and minimum of the potential dipole. Connection length is calculated by assuming that:

\begin{equation}
  L_\parallel = \frac{2a}{tan\left(\frac{\delta B}{B}\right)}
\end{equation}

where $\delta B / B$ is the poloidal magnetic field over the toroidal magnetic field.  In completely vertical field cases, this reduces to $B_z/B$.

\begin{figure}[h!]
\includegraphics[width=120mm]{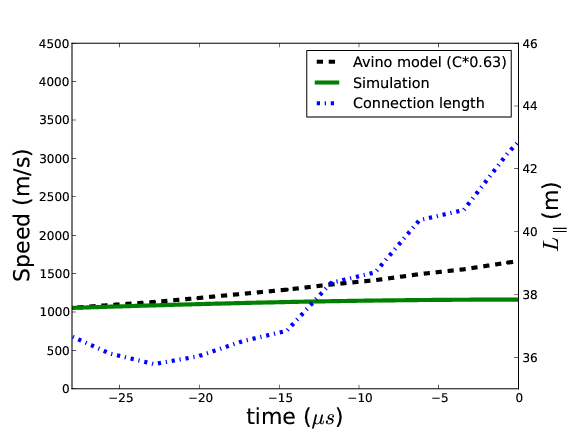}
\caption{\small{Comparison of simulated blob velocities (on a stationary background) with an analytical model~\cite{Avino2016}.  Simulated results are shown in green, the model explicitly calculated herein is shown in black, dashed, both with an adjustment to the proportionality coefficient. Here, as with~\cite{Avino2016}, t=0 is when the blob is at the null point. The connection length is shown in blue, dot-dashed.}}
\label{fig:modelcomp1}
\end{figure}

Initial simulations were performed with a stationary background plasma profile. To compare with experiment, the center of mass radial velocity was calculated at each timestep by evaluating the time derivative of the location of the center of mass. The analytical model was then plotted against the stationary background simulation shown in Figure~\ref{fig:modelcomp1}.  The proportionality coefficient $C$ is adjusted such that the calculated blob velocity coincides with our simulation 28{$\mu$}s prior to the filament arriving at the X-point.  This is also done in Reference~\cite{Avino2016}, where the proportionality constant is three times that calculated analytically.  Here, the proportionality constant is multiplied by a factor of 0.63 ($C=0.63C_{analytic}$) relative to the analytic solution. Figure~\ref{fig:modelcomp1} illustrates the simulation, the calculations based on the parameters from the simulation presented here with an adjustment to the proportionality constant, and the connection length $L_\parallel$ which was previously asserted to be the main contribution to filament acceleration.

From Figure~\ref{fig:modelcomp1} it is not clear how well the analytical model expressed in Equation~\ref{eq:Avinomodel} reproduces the data.  While the increasing connection length corresponds to an increased analytical blob velocity, the simulated filament velocity is not fully recovered, even when other factors such as $\delta n /n $ in Equation~\ref{eq:Avinomodel} are evolved.  

\begin{figure}[h!]
\includegraphics[width=120mm]{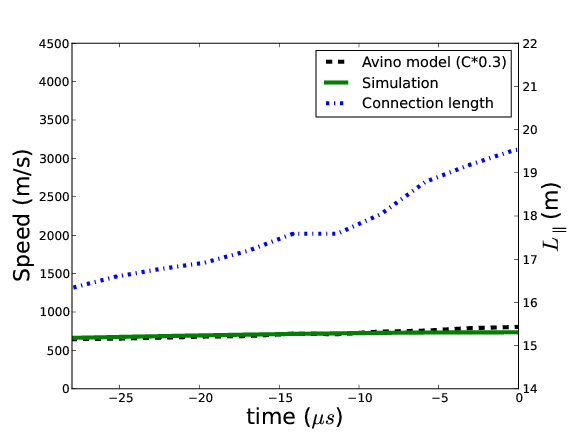}
\caption{\small{Comparison of simulated blob velocities (on a stationary background) with an analytical model~\cite{Avino2016}.  Here, the filaments are seeded farther from the X-point, at $r_0$ = -8cm to allow the dipole to fully develop.  Simulated results are shown in green and the model explicitly calculated herein is shown in black, dashed.  The connection length is shown in blue, dot-dashed.}}
\label{fig:modelcomp2}
\end{figure}

The same analysis was conducted on a filament seeded farther from the magnetic null region.  This allows the filament dipole to fully develop before encountering any effects of the X-point.  The results are shown in Figure~\ref{fig:modelcomp2}. 

This supports the hypothesis that the increasing connection length $L_\parallel$ in the region of the X-point causes an acceleration, as the model described in Reference~\cite{Avino2016} reproduces results seen in simulations.  Here, the proportionality coefficient $C$ was decreased by a factor of 3.3 relative to the analytic solution ($C=0.3C_{analytic}$). As the analytical model exhibits the same acceleration profile as shown in simulation, it is plausible to conclude that the acceleration seen in the simulations is due to the introduction of the X-point.  This acceleration, however, is smaller than that seen in experiment, and is potentially attributed to the moving background in experiment causing a larger variation in connection length.  The effects of a moving background plasma profile will be investigated in Section~\ref{sec:movingbg}. The following section investigates the effects of the X-point further by varying the seeding location and displaying the velocities of filaments throughout their entire trajectory.

\subsection{Varying the seeding location}
\begin{figure}[h]
\includegraphics[width=110mm]{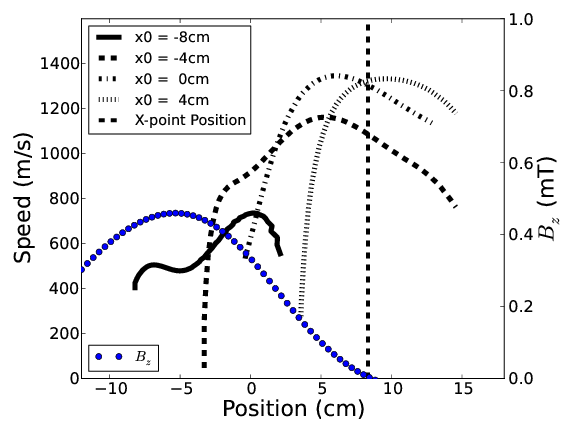}
\caption{\small{Velocity comparison of blobs seeded at various distances from the X-point.  Faster blob propagation is seen near the null region.  The vertical field strength is plotted as an indication of connection length, as $L_\parallel \sim  B_z^{-1}$.}}
\label{fig:positionscan}
\end{figure}

As stated previously, the acceleration found in experiment is much higher than that of the stationary background simulations in the region of the X-point as a moving background would introduce a stronger variation in $L_\parallel$.  However, it is still possible to determine the effect of the magnetic null region on filament propagation in stationary backgrounds by seeding blobs at various distances from the magnetic null and measuring their velocities as they approach the X-point.  The results of these simulations are shown in Figure~\ref{fig:positionscan}. 

\begin{figure}[h!]
\includegraphics[width=110mm]{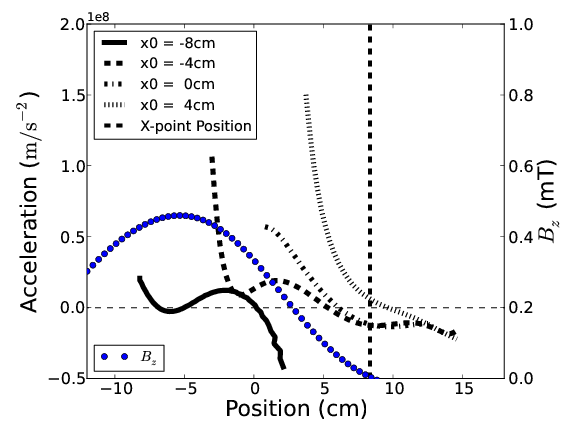}
\caption{\small{Comparison of acceleration of blobs seeded at various distances from the X-point.  The highest acceleration occurs initially, as a dipole is developing.  The vertical field strength is again shown as an indication of connection length.}}
\label{fig:blobacceleration}
\end{figure}

The acceleration of the various seeded blobs is illustrated in Figure~\ref{fig:blobacceleration}.  Filaments have a higher acceleration at the beginning of their evolution due to the developing dipole, and continue to accelerate more slowly as they approach the X-point. This supports the assertion that the magnetic null point region causes an acceleration of filaments, most likely due to the increased connection length. However, as the strongest acceleration occurs during the formation of the dipole (e.g. $\sim 1\times 10^{8}$ms$^{-2}$ for the case seeded at $x_0$ = -4cm), these results could indicate that the acceleration seen in experiment is due to the dipole forming on a moving background (which is itself approaching the null region).  This hypothesis will be further tested in Section~\ref{sec:movingbg}. 

We now show that the acceleration seen in experiment is characteristic of the initial dipole formation.  If the developing dipole were advected toward the X-point, it could appear that the magnetic null region is causing the acceleration, when in actuality the effect of the null region on the acceleration is minimal (as shown here).  To test the assertion of an advected dipole creating the acceleration profile seen in experiment, a moving background was added to the simulations, corresponding to a vertical electric field observed in experiments which created the background velocity measured in~\cite{Avino2016}.

\subsection{Constant translational background}
\label{sec:movingbg}

\begin{figure}\centering

\includegraphics[width=100mm]{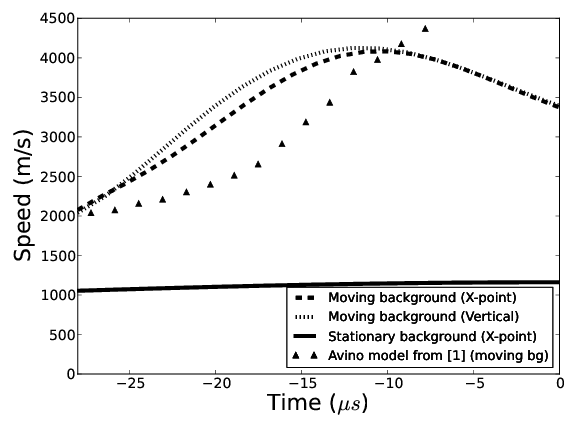}
\caption{\small{Center of mass velocity measurements from simulations of three different scenarios; stationary background X-point case (solid), moving background X-point (dashed) and vertical (dotted) fields.  The vertical field case in a moving background recovers the same characteristics as the X-point simulation and experimental measurements, indicating that the null region has little measurable effect on filament acceleration.  This assertion is also supported by the small acceleration seen in the stationary background case (solid).}}
\label{fig:simsum}

\end{figure}

To investigate if the initial dipole development causes the acceleration seen in experiment, a constant background radial plasma velocity of 2km/s was implemented in accordance with experimental measurements~\cite{Avino2016}. This was incorporated by implementing a background plasma potential profile with a constant gradient in {\bf{z}}, thereby creating a constant radial $\textbf{E} \times \textbf{B}$ motion of the plasma. Figure~\ref{fig:simsum} shows the results of three simulations.  Simulations of blobs on a stationary background plasma profile, as discussed previously, is shown as the solid line.  The dashed line indicates the velocity of blobs in a TORPEX X-point geometry with a moving background.  When  this is compared with the experimental measurements in Reference~\cite{Avino2016}, it is clear that the simulation has more closely reproduced the experimentally observed acceleration and deceleration.




Not only does this case match the velocity seen in experiment, but the average acceleration and deceleration is reproduced. There is a slight difference in the maximum velocity which could potentially be attributed to the isothermal and inviscid approximations.  Furthermore, a dipole is already present in experiment when the blob is considered coherent, which could lead to inconsistencies between the studies here and the experimental observations.  Figures~\ref{fig:modelcomp2} and~\ref{fig:positionscan}, however, would include the possibility of an already formed potential dipole, as blobs seeded further from the X-point will have completed the initial acceleration regime once they reach r=-4cm.

Figure~\ref{fig:simsum} also illustrates the calculated blob velocity using Equation~\ref{eq:Avinomodel} and the parameters from the simulation.  It is clear that the acceleration profile is not matched by the analytical model when the parameters in Equation~\ref{eq:Avinomodel} are explicitly calculated, and the initial acceleration is underestimated, indicating an additional acceleration mechanism to the increasing connection length $L_\parallel$. 

To verify that this effect is an effect of dipole formation and not the null region increasing connection length, we can overplot the velocity in a vertical magnetic field case, where no magnetic X-point is present.  The vertical field case is the typical TORPEX scenario, and has implemented via Equation~\ref{eq:biotsavart} knowing the vertical coil current and locations~\cite{FurnoPC}.  The vertical field is relatively constant and the same strength as the X-point field at the blob seeding/birth location, (r,z)=(-4cm,0cm).   The results of this test case are also shown in Figure~\ref{fig:simsum}, where the dotted line indicates the blob propagation in a vertical field case with a moving background.

Figure~\ref{fig:simsum} indicates that filaments in a vertical field have similar acceleration and velocity characteristics to those in magnetic X-point scenarios.  Additionally, the differences in velocity profiles seen in simulation lie within the experimental uncertainty~\cite{Avino2016}.  From these results it is possible to conclude that the acceleration mechanism seen in experiment is not primarily due to the increased connection length in the region of the X-point.  Instead, the moving background causes the developing dipole to propagate towards the null region as it begins to accelerate the filament relative to the background.  It should be noted that the recent experiments in magnetic null point geometries are not the first to exhibit the shown acceleration and deceleration profile.  This characteristic has been seen previously in TORPEX without poloidal magnetic nulls both with simulation~\cite{Halpern2014} and experiment~\cite{Riva2016}, both of which exhibit an initial acceleration and deceleration in the first tens of microseconds.  Additionally, the analytical model derived in~\cite{Avino2016} was also unable to explain the deceleration after $t\sim-10\mu s$ in the immediate vicinity of the X-point, which was attributed to the dissolution of the blob (despite $\delta n/n$ being considered almost constant immediately prior).  The advection of a developing dipole exhibits both an acceleration and a deceleration of the filaments on a correct timescale. As the analytical model in Figure~\ref{fig:simsum} underestimates the acceleration, the increasing connection length can be considered a minor factor in the filament acceleration.  

\section{Conclusions and future work}
\label{sec:conclusions}

We have successfully been able to model blob propagation in the X-point scenarios within the TORPEX device using a method of perturbed magnetic vector potentials.  Experimental measurements could be reproduced, however simulation results indicate that the filament acceleration seen in experiment is due to dipole formation, and not the increased connection length caused by to the introduction of an X-point. 
It has also been shown that the magnetic null region does indeed cause an acceleration of filaments in the vicinity of the X-point.  This acceleration, however, is much smaller than that of the initial dipole formation, and therefore is difficult to measure experimentally.  However, if the magnetic null were created farther from the region where the filaments are formed, it would in principle be possible to measure the acceleration due to the increased connection length in the X-point region, provided the blob dipoles were given sufficient time to form.  Future computational analysis of TORPEX configurations should look to implement a more complicated model which does not make an isothermal approximation and more accurately incorporates neutrals.    

\subsection*{Acknowledgements}  

The authors would like to thank Fabio Avino, Ivo Furno, Paolo Ricci, Christian Theiler, and Ambrogio Fasoli for their cooperation and many helpful discussions regarding this work. 
This work has been carried out within the framework of the EUROfusion Consortium and has received funding from the Euratom research and training programme 2014-2018 under grant agreement No 633053. The views and opinions expressed herein do not necessarily reflect those of the European Commission. The authors acknowledge access to the ARCHER computing service through the Plasma HEC Consortium EPSRC grant number EP/L000237/1. 

\tiny{
\bibliography{torpex}
\bibliographystyle{unsrt}}
\end{document}